\documentstyle[12pt,psfig,draft,lscape,array,amssymb,longtable]{nature-ejb-new}

\def\tip{t_i^{\prime}}
\def\tjp{t_j^{\prime}}

\def\be{\begin{equation}}
\def\ee{\end{equation}}
\def\beq{\begin{eqnarray}}
\def\eeq{\end{eqnarray}}

\def\simlt{\mathrel{\hbox{\rlap{\hbox{\lower4pt\hbox{$\sim$}}}\hbox{$<$}}}}
\def\simgt{\mathrel{\hbox{\rlap{\hbox{\lower4pt\hbox{$\sim$}}}\hbox{$>$}}}}

\def\ale{\mathrel{\hbox{\rlap{\hbox{\lower4pt\hbox{$\sim$}}}\hbox{$<$}}}}
\def\age{\mathrel{\hbox{\rlap{\hbox{\lower4pt\hbox{$\sim$}}}\hbox{$>$}}}}

\begin{document}

\title{Fast cooling synchrotron radiation in a decaying magnetic field and 
$\gamma$-ray burst emission mechanism}

\author{Z. Lucas Uhm\affiliation[1]{Kavli Institute of Astronomy and Astrophysics,
Peking University, Beijing 100871}~\affiliation[2]{Department of Physics and
Astronomy, University of Nevada Las Vegas, NV 89154, USA}, Bing
Zhang\affiliationmark[1]~\affiliationmark[2]~\affiliation[3]{Department of Astronomy,
Peking University, Beijing 100871}}

\headertitle{Fast cooling synchrotron radiation and GRB emission}
\mainauthor{Uhm \& Zhang}

%
%

\summary{Synchrotron radiation of relativistic electrons is an important radiation mechanism in many astrophysical sources. In the sources where the synchrotron cooling time scale $t_c$ is shorter than the dynamical time scale $t_{dyn}$, electrons are cooled down below the minimum injection energy. It has been believed that such ``fast cooling'' electrons have an energy distribution $dN_e /d\gamma_e \propto \gamma_e^{-2}$, and their synchrotron radiation flux density has a spectral shape\cite{sari98} $F_\nu \propto \nu^{-1/2}$. On the other hand, in a transient expanding astrophysical source, such as a gamma-ray burst (GRB), the magnetic field strength in the emission region continuously decreases with radius. Here we study such a system, and find that in a certain parameter regime, the fast cooling electrons can have a harder energy spectrum, and the standard $d N_e / d \gamma_e \propto \gamma_e^{-2}$ spectrum is achieved only in the deep fast cooling regime when $t_c \ll t_{dyn}$. We apply this new physical regime to GRBs, and suggest that the GRB prompt emission spectra whose low-energy photon index $\alpha$ has a typical value\cite{band93,preece00,zhang11,nava11} -1 could be due to synchrotron radiation in this moderately fast cooling regime. }
\maketitle

The radiation mechanism of GRBs, the
most luminous explosions in the universe, remains un-identified after 45
years since their discovery in late 1960s. 
A typical GRB prompt emission spectrum is a smoothly connected broken
power law called the ``Band function''\cite{band93}, whose typical low-
and high-energy photon spectral indices (in convention of $d N / d E_\gamma
\propto E^{\alpha}$ or $\propto E^{\beta}$) are $\alpha \sim -1$ and 
$\beta \sim -2.2$. Synchrotron radiation of
electrons accelerated in relativistic shocks has been suggested as
the leading mechanism\cite{meszaros94,daigne98}. However, for nominal
parameters, the magnetic field strength in the GRB emission region is 
strong enough so that the electrons are in the fast cooling regime
$t_c < t_{dyn}$. In this regime, it has been believed that the photon
index should be -1.5 (corresponding to $F_\nu \propto \nu^{-1/2}$)\cite{sari98}.
As a result, fast cooling synchrotron mechanism has been disfavored\cite{ghisellini00},
and some extra factors, such as slow cooling in a decaying magnetic 
field\cite{peerzhang06} or slow heating\cite{asano09} have to be introduced.

This well-known index $\alpha=-1.5$ can be derived
from a simple argument. Let us
consider a continuity equation of electrons in energy space
$ ({\partial / \partial t}){(d N_e / d\gamma_e)} + {(\partial 
    / \partial \gamma_e)} \left[ \dot{\gamma_e} {(d N_e /
   d\gamma_e)}\right] = Q(\gamma_e,t)$,
where $d N_e/d \gamma_e$ is the instantaneous electron spectrum
of the system at the epoch $t$,
and $Q(\gamma_e,t)$ is the source function above a minimum injection 
Lorentz factor $\gamma_m$ of the electrons. For synchrotron 
radiation, the electron energy loss rate is
\begin{equation}
\dot{\gamma_e} = - \frac{\sigma_T B^2 \gamma_e^2}{6\pi m_e c}\propto - \gamma_e^2 B^2,
\end{equation}
where $\sigma_T$, $m_e$, and $c$ are Thomson cross section, electron mass, and
speed of light, respectively, and $B$ is the strength of magnetic fields in the
emission region. For fast cooling, electrons are cooled rapidly to an energy 
$\gamma_c(t)$ (cooling energy) below the injection energy $\gamma_m$ at time $t$. 
In the regime $\gamma_c < \gamma_e < \gamma_m$, one has 
$Q(\gamma_e,t) = 0$. Also consider a steady state system
($\partial/\partial t = 0$), 
then one immediately gets $d N_e/d\gamma_e \propto \gamma_e^{-2}$, i.e. the electron
spectral index is $\tilde p=2$. The specific intensity of synchrotron spectrum 
would have a spectral index\cite{rybicki79} $s=(\tilde p-1)/2 = 1/2$ (with the
convention $F_\nu \propto \nu^{-s}$). The photon
spectral index (defined as $dN_\gamma/dE_\gamma \propto E_\gamma^{\alpha}$,
where $E_\gamma$ is the photon energy, and $N_\gamma$ is the photon number
flux) would then be $\alpha = -(1+s)=-1.5$.

The above argument relies on a crucial assumption of a steady state, which is
achieved when a constant $B$ is invoked. However, in a rapidly expanding source
such as a GRB, the magnetic field strength in the emission region cannot be preserved as a constant. 
In the rest frame of a conical jet, flux conservation indicates that\cite{spruit01}
the radial (poloidal) magnetic field component decreases as $B'_r \propto r^{-2}$, while the
transverse (toroidal) magnetic field component decreases as $B'_t \propto r^{-1}$. As a 
result, at a large radius from the central engine where $\gamma$-rays are
radiated, one has a toroidal-dominated magnetic field with $B' \propto r^{-1}$.
Here $r$ is the distance from the central engine. 
Considering other effects (e.g. magnetic dissipation, non-conical geometry), 
the decay law may be described by a more general form
\begin{equation}
B'(r)=B'_0 \left(\frac{r}{r_0}\right)^{-b}.
\label{B}
\end{equation}

We investigate a generic problem of electron fast cooling in a decreasing magnetic
field delineated by Eq.\ref{B}, and study the synchrotron emission spectrum. 
Targeting on interpreting the GRB prompt emission spectra, we adopt some parameters
that are relevant for GRBs. To be more generic, 
our calculation does not specify a particular energy dissipation 
mechanism or particle acceleration mechanism, and hence, can apply to a 
variety of GRB prompt emission models such as internal shocks\cite{rees94,daigne11}
and internal collision-induced magnetic reconnection and turbulence\cite{zhangyan11}. 
We consider a toy box that
contains electrons and a co-moving magnetic field $B'$, which moves 
relativistically towards the observer with a bulk Lorentz factor $\Gamma$.
The relativistic electrons are accelerated into a power-law distribution 
$Q(\gamma_e,t')=Q_0(t')(\gamma_e/\gamma_m(t'))^{-p}$ 
(for $\gamma_e > \gamma_m(t')$) of a slope $p$ 
and continuously injected into the box 
at an injection rate $R_{\rm inj}(t')
=\int_{\gamma_m}^\infty Q(\gamma_e,t') d\gamma_e$, where $t^{\prime}$ 
is the time measured in the co-moving fluid frame.
Here $R_{\rm inj}(t') \delta t'$ 
gives the number of electrons injected into the box 
during the time interval $t^{\prime}$ and $t^{\prime}+\delta t^{\prime}$. 

Electrons undergo both radiative and adiabatic cooling. In the rest frame that
is co-moving with the relativistic ejecta,
the evolution of the Lorentz 
factor $\gamma_e$ of an electron can be described by\cite{uhm12}
(noticing pressure $p$ is $\propto n_e^{4/3}$ in an adiabatically expanding
relativistic fluid)
\be
\label{eq:gamma_e}
\frac{d}{d t^{\prime}} \left(\frac{1}{\gamma_e}\right) = 
\frac{\sigma_T}{6\pi m_e c}\, {B'}^2 - 
\frac{1}{3} \left(\frac{1}{\gamma_e}\right) \frac{d \ln n_e}{d t^{\prime}}.
\ee
For a conically expanding toy box, we take the comoving electron number 
density $n_e \propto r^{-2}$, which gives $d \ln n_e = -2 d \ln r$.
We divide the injection function $Q(\gamma_e,t')$ into small divisions 
in time space $t^{\prime}$ and also in the energy space $\gamma_e$, 
and numerically follow cooling of each group of electrons 
(between $[t^{\prime},\, t^{\prime}+\delta t^{\prime}]$ and 
$[\gamma_e,\, \gamma_e+\delta \gamma_e]$) individually using Eq.\ref{eq:gamma_e}.
We then find the instantaneous global electron spectrum $d N_e/d \gamma_e$ 
of the system at any epoch.

We first consider four models with different decay indices $b$ in Eq.\ref{B}.
The ``normalization" parameter of magnetic field decay law 
is taken as $B'_0 = 30$ G at $r_0 = 10^{15}$ cm, and a constant injection rate
$R_{\rm inj} = 10^{47}~{\rm s^{-1}}$ is adopted.
Model [a] takes the unphysical parameter $b=0$, i.e. a constant co-moving magnetic 
field $B'=B'_0 = 30$ G, in order to be compared with other models. It implies
that there should be no change in the volume of the box. Thus for
this model we drop out the adiabatic cooling term from Eq.\ref{eq:gamma_e}. 
As shown in Column 1 of Fig.1 and Fig.2, this model gives the familiar electron 
spectrum $d N_e / d\gamma_e \propto \gamma_e^{-2}$ below $\gamma_m$, 
and the well-known photon spectrum $F_\nu \propto \nu^{-1/2}$ in the fast cooling 
regime. One can see that the standard fast cooling spectrum is reproduced
for a steady state system with a constant $B'$ and $R_{\rm inj}$.
Model [b] takes $b=1.0$ in Eq.\ref{B}. 
This is the case of free conical expansion with flux conservation (no significant
magnetic dissipation). As shown in Column 2 of 
Fig.1 and Fig.2, the electron spectrum and the photon spectrum
both harden with time. At 1.0 s after injecting the first
group of electrons, the global electron energy spectral index deviates
significantly from the $\tilde p=2$ nominal value, and hardens to around 
$\tilde p \sim 1$. The
corresponding photon spectrum is nearly flat ($F_\nu \propto \nu^0$),
which corresponds to a photon index $\alpha \sim -1$, the typical low-energy
photon index observed in most GRBs\cite{preece00,zhang11}.
In Columns 3 and 4 of Fig.1 and Fig.2, we present Models [c] and [d],
for which a steeper decay index $b=1.2$ and $b=1.5$ are adopted, 
respectively. They may correspond to the cases when significant magnetic dissipation 
occurs during the course of synchrotron radiation. 
As shown in Fig.2, both models also give spectra that are
consistent with the observations.

In order to understand the physical origin of such an effect, in Fig.3
we decompose the $t_{\rm obs} = 1.0$ s instantaneous electron spectrum
into the contributions of 10 injection time slices, each lasting 
for 0.1 s. For the constant $B'$ case (Fig.3a), one can see that as the 
electrons age, they tend to distribute more narrowly in logarithmic energy 
space, so that the electron number per energy bin increases. This is because in the
fast cooling regime, as time elapses, the original electrons with a wide
range of energy distribution tend to cool down to a narrow range of
cooling energy defined by the ages of the electrons in the group, which
are very close to each other at late epochs. Above $\gamma_m$,
the electron energy density distribution remains unchanged with time, 
since it is always determined by the same injection rate and cooling rate.

The cases of $B'$ decay show a more complicated behavior. The distribution
of each group of electrons still shrinks as the group ages. However, 
since at early epochs
the magnetic field was stronger, it had a stronger cooling effect 
so that for a same injection time duration (0.1 s), initially it had a wider spread 
in energy at a given age (which can be noticed by comparing the 0.1 s electron spectrum 
for Model [a] and [b] in Fig.1). The later injected electrons are cooled in a weaker
$B'$ field, so that their intial spread is narrower. After the same shrinking effect
due to cooling pile-up, 
the groups injected in earlier time slices have a wider electron distribution 
than the constant $B'$ case. Also the electron spectrum above the injection
energy, even though possessing a same spectral index, has a normalization
increasing with time due to progressively less cooling in a progressively 
weaker magnetic field. These complicated effects all work in the direction
to harden the spectral index, as seen in Figures 3b, 3c and 3d.  
For a steeper $B'$-decay index (e.g. $b=1.2$ and $b=1.5$), the late-time
injection occurs in an even weaker magnetic field, so that slow cooling is
possible. This results in accumulation of electrons around the 
minimum injection energy $\gamma_m$,
so that a sharper break in the electron energy distribution is achieved.

The model predicts that the low-energy spectrum below the 
injection frequency $\nu_m$ is curved, due to the complicated cooling effect as
delineated in Fig.3. Most GRB detectors have a narrow band pass so that below
the peak energy (typically a few hundred keV), there are at most 2 decades
in energy. 
Nonetheless, in the detector band pass, the observed spectra 
are usually fit by a Band function, with the low energy spectral index 
$\alpha \sim -1$. In most situations, time resolved spectral analyses are carried out
with a time bin in seconds\cite{zhang11}. This is the typical time scale of 
the slow variability component in most GRB light curves\cite{gao12}.
We therefore focus on the 1 s and 3 s model spectra. We
truncate these spectra in a narrow band (5 keV - 5 MeV) and compare 
them with the empirical Band function fits (Fig.4). One can 
observe that most of our model spectra are consistent with 
the Band function with the correct low-energy spectral indices.

Outside the band pass, our model predicts an asymptotic value of the low-energy
electron energy spectral index of $\tilde p_{a}=(6b-4)/(6b-1)$, which is 2/5 for $b=1$. 
This is seen in the numerical results of the models (lower panels in Fig.1), 
and can be derived analytically (see Appendix).
According to the simple relationship $s=(\tilde p-1)/2$, 
one gets $s_a = -3/(12b-2)$,
which is -0.3 for $b=1$ (or $F_\nu \propto \nu^{0.3}$). 
In reality, due to the contribution of the 1/3 segment of the individual 
electron spectrum, which becomes significant when $\tilde p$ approaches 1/3 from 
above, the asymptotic photon spectrum limit is softened. In this case,
$s$ is about -0.2. This corresponds to a photon index of -0.8, which is
much harder than the nominal value -1.5. 

Besides the decay index $b$ as discussed above, the value of low-energy photon index 
$\alpha$ also depends on several other factors: the 
``normalization'' parameter $B'_0$ at $r_0 = 10^{15}$ cm, 
the time history of electron injection, and the bulk Lorentz factor $\Gamma$. 
In order to see how different parameters affect the predicted $\alpha$ values,
we have carried out more calculations by varying these parameters.

The ``normalization'' factor $B'_0$ is essential in defining the strength of magnetic 
fields seen by an electron during the cooling process since injected. 
So far, we have adopted the value $B'_0 = 30$ G. In the following, 
we calculate the cases for $B'_0 = 10, 100, 300$ G for $b=1$ and constant injection
rate (Models [e], [f], and [g], respectively). To compare with model [b]
($B'_0=30$ G and $b=1$), we keep 
$\Gamma = 300$ fixed and the product of $\gamma_{m}^2 B'$ as a constant to assure a 
same observed $E_p$. We then repeat the calculations as described above
and perform the Band-function fits to the model spectra. The resulting Band-function 
parameters are presented in Table \ref{tab:B0}. One can see that an $\alpha$ value 
ranging from $\sim -1$ to $-1.5$ are obtained. The general trend is that a lower $B'$ 
tends to give rise to a harder $\alpha$ value.

  \begin{table}
    \caption[]
      {Spectral parameters of Models [e], [b], [f], and [g] (constant injection rate)}
\begin{center}
    \addtolength\tabcolsep{2pt}
      \begin{tabular}{@{}c@{\hspace{25pt}}ccccc@{}}
        \hline \hline
        Model & $B'_0$ (G)  & $t_{obs}$ (s) & $\alpha$ & $\beta$ & $E_p$ (keV)\\
        \hline
        {\rm [e]} & 10 & 1.0 & -1.03 & -2.13 & 480 \\
	{\rm [e]} & 10 & 3.0 & -1.03 & -2.10 & 220 \\
        {\rm [b]} & 30 & 1.0 & -1.22 & -2.26 & 490 \\
	{\rm [b]} & 30 & 3.0 & -1.17 & -2.26 & 220 \\
	{\rm [f]} & 100 & 1.0 & -1.42 & -2.33 & 590 \\
	{\rm [f]} & 100 & 3.0 & -1.39 & -2.35 & 240 \\
	{\rm [g]} & 300 & 1.0 & -1.50 & -2.34 & 650 \\
	{\rm [g]} & 300 & 3.0 & -1.50 & -2.37 & 250
      \end{tabular}
\end{center}
\label{tab:B0}
  \end{table}

The light curve of GRBs show erratic variability, and can be de-composed as the superposition 
of many ``pulses''. The decay phase of a pulse is usually controlled by the high-latitude 
``curvature'' effect \cite{kumar00}, so the observed spectral indices are mostly defined by 
the rising phase of a pulse. An increase in the injection rate gives more weight to electrons 
that are injected later, which tend to harden the spectrum. We test how the injection history 
during the rising phase affects $\alpha$. First, we introduce a linear increase of the injection 
rate for $B'_0=10,30,100,300$ G, respectively (with $b=1$) and name the models as [e1], [b1], 
[f1], and [g1], respectively. The fitted spectral parameters of these models are presented in 
Table \ref{tab:linear}. One can see that by introducing a rise of injection rate with time, 
the resulting $\alpha$ values are systematically harder. For the four models discussed, the 
$\alpha$ value ranges from -0.92 to -1.48.

  \begin{table}
    \caption[]
      {Spectral parameters of Models [e1], [b1], [f1], and [g1] (linear increase of injection rate)}
\begin{center}
    \addtolength\tabcolsep{2pt}
      \begin{tabular}{@{}c@{\hspace{25pt}}ccccc@{}}
        \hline \hline
        Model & $B'_0$ (G)  & $t_{obs}$ (s) & $\alpha$ & $\beta$ & $E_p$ (keV)\\
        \hline
        {\rm [e1]} & 10 & 1.0 & -0.92 & -2.09 & 460 \\
	{\rm [e1]} & 10 & 3.0 & -0.92 & -2.07 & 220 \\
        {\rm [b1]} & 30 & 1.0 & -1.13 & -2.24 & 460 \\
	{\rm [b1]} & 30 & 3.0 & -1.09 & -2.23 & 210 \\
	{\rm [f1]} & 100 & 1.0 & -1.37 & -2.32 & 520 \\
	{\rm [f1]} & 100 & 3.0 & -1.34 & -2.33 & 220 \\
	{\rm [g1]} & 300 & 1.0 & -1.48 & -2.33 & 610 \\
	{\rm [g1]} & 300 & 3.0 & -1.48 & -2.35 & 240
      \end{tabular}
\end{center}
\label{tab:linear}
  \end{table}

The rising phase may be steeper than a linear increase with time. We next test the effect of 
different rising profiles on $\alpha$. We fix $B'_0=10$ G in order to check how hard a spectrum 
one may get. Considering the injection rate $Q(t') \propto {t'}^q$, we calculate the cases for 
$q=0,1,2,3$ (Models [e], [e1], [e2], and [e3], respectively). Table \ref{tab:t} shows the 
spectral parameters of these models. One can see that $\alpha$ hardens as $q$ increases (a 
more rapid increase). For these four models, the $\alpha$ value is in the range between 
-0.82 and -1.03.

  \begin{table}
    \caption[]
      {Spectral parameters of Models [e], [e1], [e2], and [e3] ($B'_0 = 10$ G)}
\begin{center}
    \addtolength\tabcolsep{2pt}
      \begin{tabular}{@{}c@{\hspace{25pt}}ccccc@{}}
        \hline \hline
        Model & $q$  & $t_{obs}$ (s) & $\alpha$ & $\beta$ & $E_p$ (keV)\\
        \hline
        {\rm [e]} & 0 & 1.0 & -1.03 & -2.13 & 480 \\
	{\rm [e]} & 0 & 3.0 & -1.03 & -2.10 & 220 \\
        {\rm [e1]} & 1 & 1.0 & -0.92 & -2.09 & 460 \\
	{\rm [e1]} & 1 & 3.0 & -0.92 & -2.07 & 220 \\
	{\rm [e2]} & 2 & 1.0 & -0.87 & -2.06 & 470 \\
	{\rm [e2]} & 2 & 3.0 & -0.87 & -2.03 & 220 \\
	{\rm [e3]} & 3 & 1.0 & -0.82 & -2.04 & 470 \\
	{\rm [e3]} & 3 & 3.0 & -0.82 & -2.01 & 220
      \end{tabular}
\end{center}
\label{tab:t}
  \end{table}

This model predicts a hard-to-soft evolution of the peak energy
$E_p$ during a broad pulse. This is consistent with the observational trends of a large
fraction of GRBs \cite{lu12}. According to Fig.1 and Fig.2, the electron spectrum also
tends to harden with time. So is the $\alpha$ value.  This model therefore predicts that 
for a broad pulse in a GRB, during the very early epochs, the $\alpha$ value would harden 
with time. If the $\alpha$ value of a GRB is already very hard from the very beginning, 
then the above mentioned $\alpha$ evolution is no longer significant, even though 
electron spectrum continues to harden with time. This is because the contributions 
from the 1/3 spectral segment for individual electrons become more important. 

Two caveats to apply this model to interpret GRB prompt emission should be noticed.
First, observations showed that a growing sample of GRBs have a quasi-thermal component
superposed on the Band component \cite{guiriec11,axelsson12,guiriec13}. While the Band
component is likely of a synchrotron origin \cite{burgess11,zhang12,veres12}, the
quasi-thermal component is widely interpreted as emission from the GRB photosphere
\cite{meszaros00,peer06,beloborodov10,lazzati13}, the relative strength of which with
respect to the synchrotron component depends on the composition of the GRB ejecta,
and could be dominant if the ejecta is a matter-dominated fireball. Since the
hardest $\alpha$ value we get is about -0.8, an observed $\alpha$ harder than this
value would be the evidence of a dominant photosphere component \cite{meszaros00}. 
Second, in order to interpret the typical $\alpha \sim -1$ in our model, some
requirements to the parameters are needed. 
A plausible scenario to satisfy these parameter constraints would be magnetic dissipation 
models that invoke a large dissipation radius, such as the internal collision-induced 
magnetic reconnection and turbulence (ICMART) model\cite{zhangyan11}. Due to the large 
emission radius $R \simgt 10^{15}$ cm, this model allows  seconds-duration 
broad pulses as fundamental radiation units, during which particles are continuously 
accelerated. Due to a moderately high magnetization parameter $\sigma$ in the emission 
region, the minimum injected electron Lorentz factor $\gamma_m \sim 10^5$ can be 
achieved, since a small amount of electrons share a similar amount of dissipated 
energy. One potential difficulty is that there is 
a preferred range of $B'_0$ (10 - 300 G). The magnetization parameter 
\begin{equation}
 \sigma = 2.4 \times 10^{-4}~{\left(\frac{\Gamma}{300}\right)}^2
~{\left(\frac{B'}{30~{\rm G}}\right)}^2 {\left(\frac{R}{10^{15}~{\rm cm}}\right)}^2
~L^{-1}_{52}
\end{equation}
is required to be in the range of $2.7 \times 10^{-5} - 2.4\times 10^{-2}$ for 
$\Gamma=300$, $R=10^{15}$ cm and $L=10^{52}~{\rm erg~s^{-1}}$, which is relatively
low. Within the ICMART scenario, the 
electrons likely radiate in the outflow region of a reconnection layer, 
in which magnetic fields are largely dissipated. One therefore expects a 
relatively low $B'_0$ (and hence, low $\sigma$) as 
compared with the un-dissipated regions in the outflow. 
Nonetheless, detailed studies of magnetic reconnection and particle acceleration
processes are needed to address whether the $B'_0$ range demanded by the model
could be achieved.

The new physics in the moderately fast cooling regime discussed in this 
paper would find applications in many other astrophysical systems invoking 
jets and explosions, such as active galactic nuclei, galactic ``micro-quasars''
in X-ray binaries, as well as jets from tidal disruption of stars by supermassive
black holes. Within the GRB context, it also finds application in the afterglow
phase where electrons never enter a deep fast-cooling regime. Further investigations
of this physical processes in other astrophysical environment are called for.




{\bf References and Notes}

\begin{acknowledge}
This work is partially supported by the China Postdoctoral Science Foundation
through Grant No. 2013M540813, National Basic Research Program (``973''
Program) of China under Grant No. 2014CB845800, and by National Science 
Foundation under grant AST-0908362. 
We thank helpful discussion with following colleagues: Andrei Beloborodov, 
Frederic Daigne, Gabriel Ghisellini, Kunihito Ioka, Pawan Kumar, Davide Lazzati, Zhuo Li, 
Peter M\'esz\'aros, Kohta Murase, Robert Mochkovitch, and Asaf Pe'er.
\end{acknowledge}

\bigskip
\bigskip
\centerline{\bf Appendix: Asymptotic Values}

The asymptotic low-energy spectral index can be derived analytically from
Eq.3.
Assuming a constant Lorentz factor $\Gamma$ (which is relevant for GRB prompt
emission), one has $r = c\, t^{\prime} \Gamma$. 
We first solve a simpler equation by dropping out
the adiabatic term, i.e.
\label{eq:no_adia}
\be
\frac{d}{d t^{\prime}} \left(\frac{1}{\gamma_e}\right) = a\, {t^{\prime}}^{-2b},
\ee
where 
\be
a \equiv \frac{\sigma_T}{6\pi m_e c}\, {B'_0}^2\, (c\Gamma/r_0)^{-2b}.
\ee
We then find the solution of electron Lorentz factor at any time $\tjp$ ($> \tip$)
\be
\gamma_e(\tjp)=\left[ \frac{1}{\gamma_e(\tip)}+
\frac{a}{1-2b} \left\{ {\tjp}^{1-2b}-{\tip}^{1-2b} \right\} \right]^{-1},
\ee
where $\gamma_e(\tip)$ is the electron Lorentz factor 
at an initial time $\tip$. 
For $b>1/2$, $\tjp \gg \tip$, and $\gamma_e(\tjp) \ll \gamma_e(\tip)$, this solution gives 
$\gamma_e(\tjp) \propto {\tip}^{2b-1}$. We then get
\be 
\delta \gamma_e(\tjp) \propto {\tip}^{2b-2}\, \delta \tip 
\propto \left[ \gamma_e(\tjp) \right]^{\frac{2b-2}{2b-1}}\, \delta \tip.
\ee
For a constant injection rate $R_{\rm inj}$, we have $\delta N_e \propto \delta \tip$. 
Thus, we have an asymptotic behavior of the global electron spectrum as follows.
\be
\frac{\delta N_e}{\delta \gamma_e(\tjp)} \propto \left[ \gamma_e(\tjp) \right]^{-\frac{2b-2}{2b-1}}.
\ee

Now we consider the full Eq.3 that includes the adiabatic term.
For $B'(r)=B'_0\, (r/r_0)^{-b}$ and $r = c\, t^{\prime} \Gamma$, 
Equation 3 can be written as
\be
\label{eq:adia}
\frac{d}{d t^{\prime}} \left(\frac{1}{\gamma_e}\right) = a\, {t^{\prime}}^{-2b}+
\frac{2}{3t^{\prime}} \left(\frac{1}{\gamma_e}\right).
\ee
This equation has an analytic solution
\be
\gamma_e(t^{\prime})={t^{\prime}}^{-2/3} \left[\frac{3a}{1-6b}\, {t^{\prime}}^{(1-6b)/3}+C \right]^{-1},
\ee
where $C$ is the integration constant of the differential equation, to be determined 
by the initial condition; $\gamma_e(\tip)$ at time $\tip$. 
The electron's Lorentz factor $\gamma_e(\tjp)$ at a later time $\tjp$ is found to be 
\be
\gamma_e(\tjp)={\tjp}^{-2/3} \left[ \frac{{\tip}^{-2/3}}{\gamma_e(\tip)} +
\frac{3a}{1-6b} \left\{ {\tjp}^{(1-6b)/3}-{\tip}^{(1-6b)/3} \right\} \right]^{-1}.
\ee
For $b>1/6$, $\tjp \gg \tip$, and ${\tjp}^{2/3} \gamma_e(\tjp) \ll {\tip}^{2/3} \gamma_e(\tip)$, 
this solution gives 
$\gamma_e(\tjp) \propto {\tjp}^{-2/3}\, {\tip}^{(6b-1)/3}$. 
A variation in $\gamma_e(\tjp)$ results only from $\delta \tip$ for the instantaneous 
(i.e., at a fixed time $\tjp$) global electron spectrum; 
$\delta \gamma_e(\tjp) \propto {\tjp}^{-2/3}\, {\tip}^{(6b-4)/3}\, \delta \tip$. 
This gives the asymptotic behavior of the global electron spectrum
\be
\frac{\delta N_e}{\delta \gamma_e(\tjp)} \propto 
{\tjp}^{\frac{2}{6b-1}} \left[ \gamma_e(\tjp) \right]^{-\frac{6b-4}{6b-1}},
\ee
where we have again assumed a constant injection rate $R_{\rm inj}$.
Therefore we have the asymptotic low-energy electron spectral index
$\tilde p_a = (6b-4)/(6b-1)$.

\clearpage

\begin{figure}
\begin{center}
\psfig{file=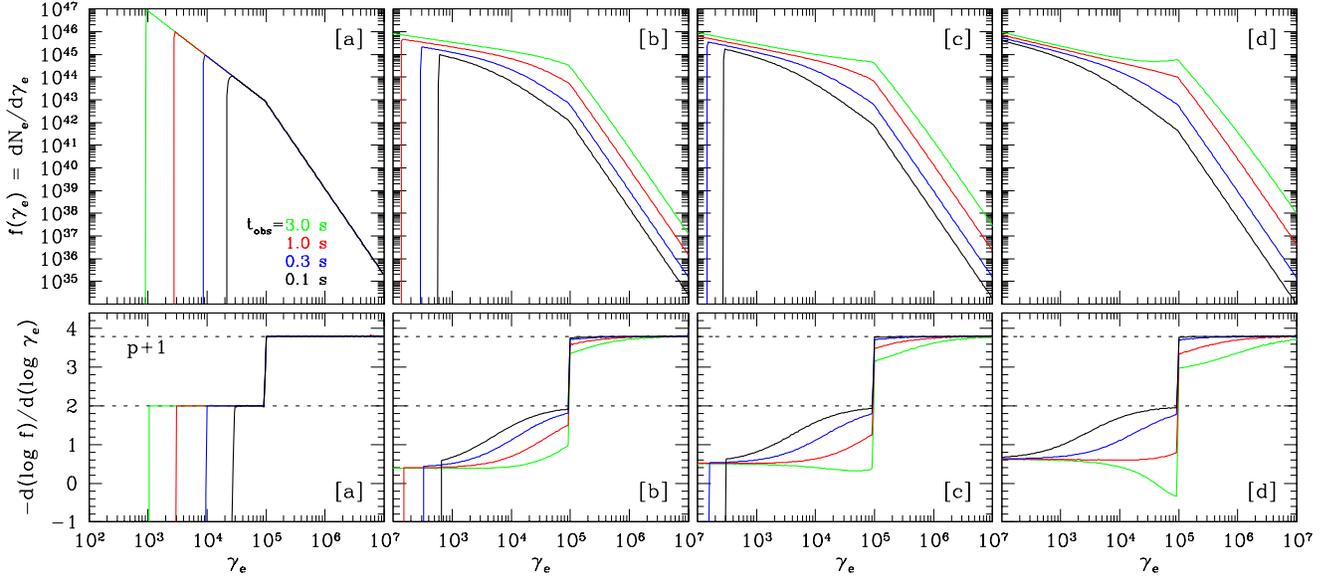}
\caption{
The co-moving frame fast cooling electron energy spectrum evolution as a function 
of time $t'$. The injected electrons have a power law distribution
$Q(\gamma_e,t') =Q_0 (t') (\gamma_e /\gamma_m)^{-p}$ above a minimum injection
Lorentz factor $\gamma_m=10^5$, with a power law index
$p=2.8$ for $\gamma_e > \gamma_m$. We take a co-moving magnetic field
$B'=B'_0 (r/r_0)^{-b}$ with $B'_0 = 30$ G and $r_0 = 10^{15}$ cm. Four models
are investigated: [a] $b=0$; [b] $b=1$; [c] $b=1.2$; and [d] $b=1.5$.
For all the models, a constant injection rate 
$R_{\rm inj} = \int_{\gamma_m}^\infty Q(\gamma_e,t') d\gamma_e=10^{47}~\mbox{s}^{-1}$
is adopted, with both $Q_0$ and $\gamma_m$ as constants.
The electron injection into the box begins at $r=10^{14}$ cm.
The ejecta is assumed to be moving towards the observer
with a Lorentz factor $\Gamma=300$, and the burst is assumed at a cosmological 
redshift $z=1$. For each model, the instantaneous electron spectra at four 
different epochs since the beginning of electron acceleration are calculated. 
The four epochs in the observer's frame are: 0.1 s (black), 0.3 s (blue),
1.0 s (red), and 3.0 s (green). For each epoch, the sharp cutoff at low
energies corresponds to the ``cooling energy'' of the system,
which is defined by the strength of the magnetic fields and the age of the electrons.
Given the same $B'_0$ field at $r=10^{15}$ cm, the $B'$ field is stronger 
at earlier epochs, so that electrons undergo more significant cooling
initially. One can see that the cooling energy is systematically lower than
that of the constant $B'$ case (Model [a]) as $b$ steepens (Models [b], [c]
and [d]). 
The lower panel of each model shows the local
electron spectral index as a function of electron energy $\gamma_e$. 
For Model [a] (constant magnetic field), the electron spectrum
shows the well-known broken power law, with the spectral indices $p+1$ and 2
above and below the injection energy, respectively. For other models, even
though the index above $\gamma_m$ remains the same,
the index below $\gamma_m$ is much harder.
At later epochs (e.g. 3 s spectra), the 
index approaches an asymptotic value $\tilde p_a=(6b-4)/(6b-1)$.
}
\label{fig:f1}
\end{center}      
\end{figure}

\begin{figure}
\begin{center}
\psfig{file=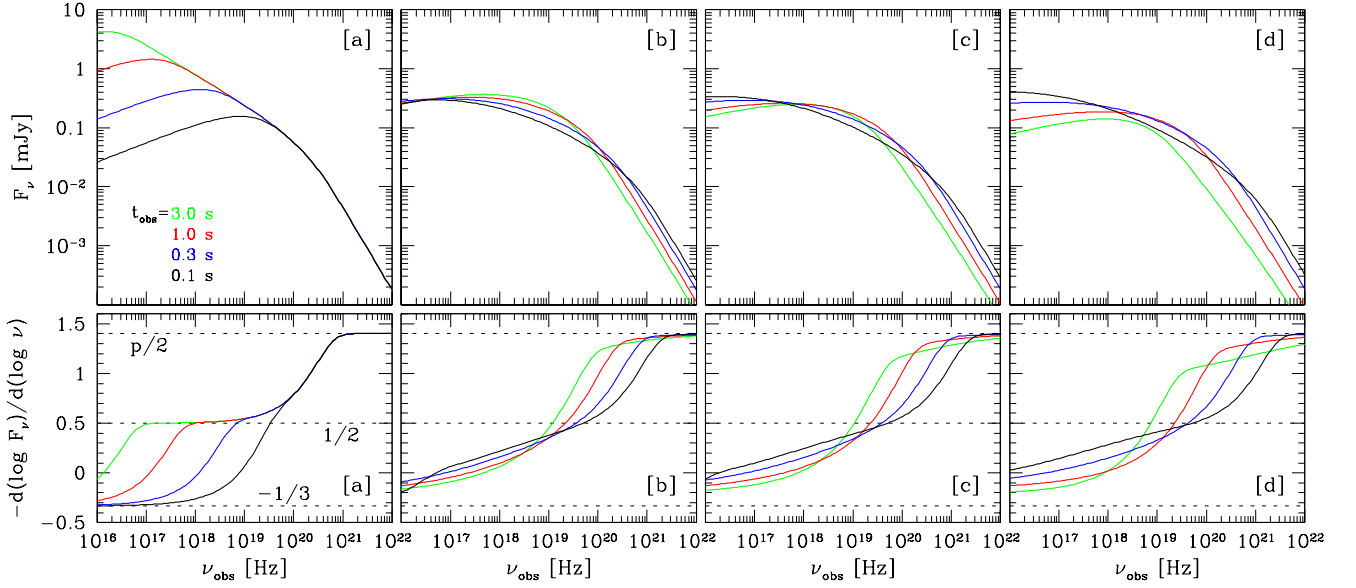}
\caption{
The synchrotron emission flux-density ($F_\nu$) spectra of electrons with energy 
distribution presented
in Figure 1. The full synchrotron spectrum of each electron\cite{rybicki79}
is taken into account. The observed spectra are calculated by considering the Lorentz
blueshift and cosmological redshift. 
While the constant $B'$ case (Model [a]) gives rise to the familiar $F_\nu
\propto \nu^{-1/2}$ spectrum, the decaying $B'$ cases (Models [b], [c] and [d])
all give rise to much harder spectrum below the injection break $\nu_m$. For the 
spectra in the seconds time scale (1 s - red; 3 s - green), the low-energy
spectral index is nearly flat, consistent with the typical observed photon index
-1. Lower panels show local spectral slopes as a function of observed frequency.
The energy peak $E_p$ corresponds to the transition break towards the $p/2$ index.
So a clear hard-to-soft evolution of $E_p$ is predicted, which is consistent with
the data of most broad pulses observed in GRBs\cite{lu12}.
 } 
\label{fig:f2}
\end{center}      
\end{figure}

\begin{figure}
\begin{center}
\psfig{file=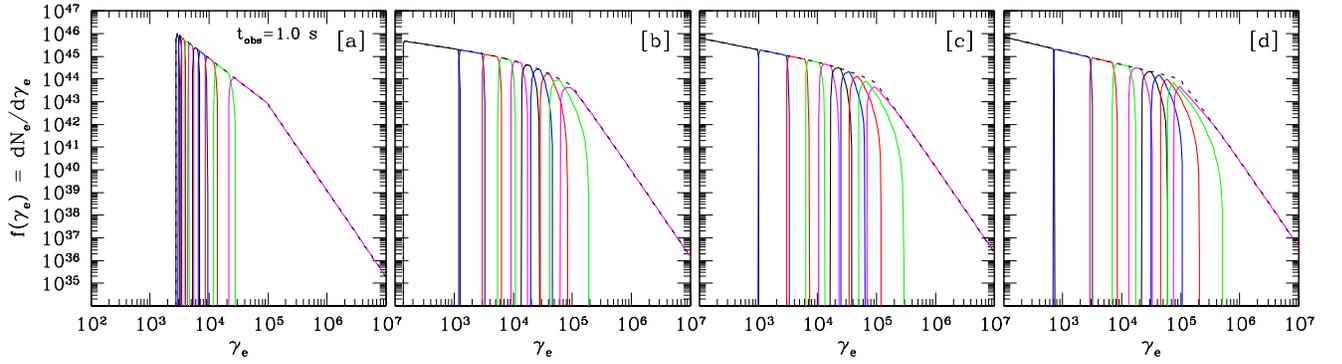}
\caption{
Decomposition of electron spectrum at 1 s in the observer's frame. 
In order to see the contributions of electrons injected at different 
epochs, the electrons are grouped into 10 slices in injection time, each 
with a duration of 0.1 s. The contributions of each group to the
instantaneous electron spectrum at 1 s are marked in different
colors. Older groups are cooled down further towards lower energies,
so from left to right, curves with different colors denote the 
electron energy distribution of the electron groups injected from 
progressively later epochs, with a 0.1 s time step. Dashed curves 
are the summed total of all electrons.
} 
\label{fig:f3}
\end{center}      
\end{figure}

\begin{figure}
\begin{center}
\psfig{file=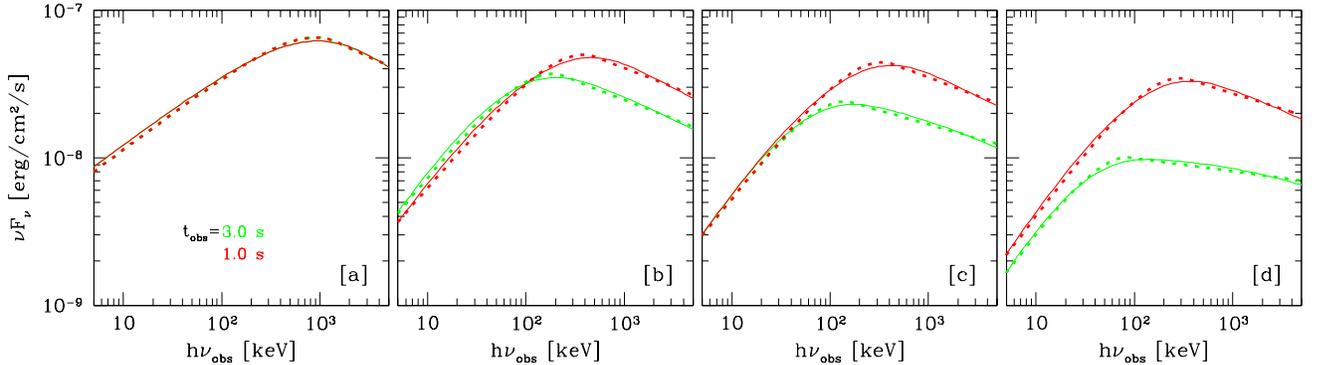}
\caption{
A comparison of our 1 s and 3 s model spectra (solid) with the empirical
Band function fits (dashed) for all four models in a narrower band pass
from 5 keV to 5 MeV. The energy spectra ($\nu F_\nu$) are presented to 
show clear peak energy ($E_p$) in the spectra. It is seen that the model 
spectra can mimic the
Band function spectra well. The plotted Band function parameters are
the following:
Model [a]: $\alpha=-1.5$, $\beta=-2.3$, $E_0 = 1800$ keV for both 1 s and 3 s;
Model [b]: $\alpha=-1.22$, $\beta=-2.26$, $E_0=490$ keV for 1 s, and
$\alpha=-1.17$, $\beta=-2.26$, $E_0=220$ keV for 3 s;
Model [c]: $\alpha=-1.16$, $\beta=-2.25$, $E_0=400$ keV for 1 s, and
$\alpha=-1.12$, $\beta=-2.19$, $E_0=160$ keV for 3 s;
Model [d]: $\alpha=-1.1$, $\beta=-2.21$, $E_0=320$ keV for 1 s, and
$\alpha=-1.05$, $\beta=-2.09$, $E_0=90$ keV for 3 s.
} 
\label{fig:f3}
\end{center}      
\end{figure}

\end{document}